# Recently predicted ternary boride $Hf_3PB_4$: Insights into the physical properties of this hardest possible boride MAX phase

M. A. Ali[a,1], M. M. Hossain[a], A. K. M. A. Islam[b,c], S. H. Naqib[c,2]

[a]Department of Physics, Chittagong University of Engineering and Technology (CUET), Chattogram4349, Bangladesh
[b]Department of Electrical and Electronic Engineering, International Islamic University Chittagong, Kumira, Chittagong, 4318, Bangladesh
[c]Department of Physics, University of Rajshahi, Rajshahi-6205, Bangladesh

## ABSTRACT

In this work, we have explored via first-principles study of mechanical properties including Vickers hardness and mechanical anisotropy, electronic charge density distribution, Fermi surface, thermodynamic and optical properties of the recently predicted thermodynamically stable MAX phase boride $Hf_3PB_4$ for the first time. The calculated lattice constants of the optimized cell are consistent with those found by the predicted data available. Mechanical properties characterized by parameters such as $C_{44}$, $B$ (bulk modulus), $G$ (shear modulus), $Y$ (Young's modulus), $H_{macro}$ (macro-hardness) and $H_{micro}$ (micro-hardness) of $Hf_3PB_4$ boride are compared with those of existing 312 and 413 MAX phases. Moreover, the calculated values of $C_{44}$, $B$, $G$, $Y$, $H_{macro}$ and $H_{micro}$ are also compared with ninety one (91) 211 MAX phase carbides, nitrides and borides. None of the MAX compounds synthesized so far has higher $H_{macro}$ and/or $H_{micro}$ than that of the predicted $Hf_3PB_4$ nanolaminate. Calculations of stiffness constants ($C_{ij}$) indicate that $Hf_3PB_4$ is mechanically stable. The extraordinarily high values of elastic moduli and hardness parameters are explained with the use of density of states (DOS) and charge density mapping (CDM). The high stiffness of $Hf_3PB_4$ arises because of the additional B atoms which results in the strong B–B covalent bonds in the crystal. The band structure and DOS calculations are used to confirm the metallic properties with dominant contribution from the Hf-5*d* states to the electronic states around the Fermi level. The technologically important thermal parameters such Debye temperature, minimum thermal conductivity, Grüneisen parameter and melting temperature of $Hf_3PB_4$ are calculated. It has been found that the estimated melting temperature of $Hf_3PB_4$ is also the highest among all the MAX phase nanolaminates. The important optical constants are calculated and analyzed in detail and their relevance to possible applications in the optoelectronic sectors is discussed. Our study reveals that $Hf_3PB_4$ has the potential to be the hardest known MAX phase based on the values of $C_{44}$, $H_{macro}$ and $H_{micro}$.

Corresponding authors: [1]ashrafphy31@cuet.ac.bd; [2]salehnaqib@yahoo.com

## 1. Introduction

The chemical formula $M_{n+1}AX_n$, n =1, 2 or 3, generally represents a large class of materials; layered ternary carbides, nitrides and borides. The M, A and X in the $M_{n+1}AX_n$ stand for early transition metals; an element belonging to groups 12-16 and either C, N or B, respectively. The carbides and nitrides are well known since 1990s after the reporting by Barsoum et al. [1,2]. Though, these materials were brought to front by Nowotny et al. [3–6] in the 1960s. The MAX phase borides are the recent extension of the X elements [7]. The interest from the scientific community has been refreshed after reporting [1,2] the exciting combination of the properties of the MAX phases. For certain aspects, they can be considered as ceramic (for practical applications) because of their properties that are normally common to ceramic materials such as being lightweight, elastically rigid, and having oxidation and corrosion resistance at high temperature. Their expansion coefficient is low as well as they possesses high strength at high temperature. Whereas, the good electrical and thermal conductivity, soft nature with machinability, thermal shock resistance etc. are importance features of MAX phases that demonstrate their metallic nature for practical applications [8–13]. Therefore, the term metallic-ceramics could be an appropriate to define the MAX phases. In recent years, study of MAX phase materials has become an important sub-class of research in materials science [14]. To predict the new compound belonging to the MAX phase family, major attempts have been devoted to the discovery of solid solutions (by mixing of M, M′ and A, A′; M, M′ = Ti, Zr, Hf, Ta, ….; A, A′ = Al, Ga, Si, Ge, P, …) [15–24], extending the M and/or A elements from the periodic table [25–34]. There are also some reports on extension of MAX phase unlike of conventional MAX phases such as $Mo_2Ga_2C$ [35,36], 321 MAX phases [37] and atomically layered and ordered rare-earth i-MAX phases[38,39].

Comparatively fewer attempts have been made to tune the X elements. Replacing of C/N by boron (B) atoms may open the door to extend the MAX phase family owing to the motivating physical and chemical properties of B and its boride compounds [40]. Recently, the boride MAX phases have been enlisted as the members of the exciting MAX phases and have been able to draw significant attentions from the scientific forum [7,41–46]. The hypothetical $M_2AlB$ (M = Sc, Ti, Cr, Zr, Nb, Mo, Hf, or Ta) MAX phase borides have been investigated by Khazaei et al. [7]. The electronic and lattice dynamical properties of $Ti_2SiB$ have been investigated by Gencer et al. [41]. Surucu et al. [42] have predicted $M_2AlB$ (M = V, Nb, Ta) MAX by confirming their dynamical stability. The $M_2SB$ (M = Zr, Hf and Nb) borides have been synthesized by Rackl et al. [43,44]. The soft MAX phase ($V_2AlB$) has been predicted by Chakraborty et al. [45]. G. Surucu [46] has performed theoretical investigation of structural, electronic, anisotropic elastic, and lattice dynamical properties of MAX phases borides [$M_2AB$ (M = Ti, Zr, Hf; A = Al, Ga, In)]. We have also performed a systematic investigation on the synthesized MAX phase borides $M_2SB$ (M = Zr, Hf andNb) [47].

Recently, Miao et al. [48] have predicted some thermodynamically stable boron based MAX phases. The prediction included a new type of thermodynamically stable ternary layered boride, $Hf_3PB_4$, defined as 314 MAX phase boride which is crystallized in the sub-space group ($P\bar{6}m2$; No. 187) of conventional MAX phase similar to that of the recently discovered ternary layered hexagonal MAX boride $Ti_2InB_2$ ($P\bar{6}m2$; No. 187) [49]. The layered ternary borides MAB phases already proposed by Ade and Hillebrecht [50] in 2015 which are further synthesized by Kota et al.[51] who explored their structure, chemical bonding as well as other characteristics properties. The crystal structure of MAB phases is different from MAX phases as the MAB compounds crystallize in the orthorhombic structure [50,52]. The M-B and B-B bonding contributed strongly in layered feature which results in significant structural stability. Wang et al. [49] have synthesized the $Ti_2InB_2$ ternary boride which crystallizes in hexagonal structure (space group:$P\bar{6}m2$, No. 187) and possesses the similar characteristics of conventional C/N containing 212 MAX phases.

Although the predicted $Hf_3PB_4$ has the same space group as $Ti_2InB_2$, it shows a different structure type with thicker M-B layers (with 5 layers of M and B) than those in $Ti_2InB_2$. Miao et al.[48] have checked the stability of $Hf_3PB_4$ and investigated the electronic properties such as band structure, density of states and electron localization function. All other physicochemical properties of the $Hf_3PB_4$ compound remained unexplored. For both MAX and MAB phase compounds, study of mechanical properties is of prime interest to predict the suitability for engineering applications. In the thermodynamically stable $Hf_3PB_4$ structure, two extra B atoms reside at X position of conventional MAX phase resulting in extra strong B-B bonding, that leads to a possibility of $Hf_3PB_4$ being the hardest among the MAX phases known so far. Therefore, a detailed study of hardness of $Hf_3PB_4$ carries a significant technological and scientific interest. Understanding of the thermal properties is prerequisite to forecast possible applications at high temperatures. Therefore, with the intention of obtain a systematic and deep understanding of the predicted $Hf_3PB_4$ compound, we have performed a thorough first-principles study of mechanical properties including Vickers hardness and mechanical anisotropy, electronic charge density distribution, Fermi surface topography, thermodynamic and optical properties for the first time. Besides, to predict and validate that $Hf_3PB_4$ is possibly the hardest among the MAX phases known so far, we have calculated the $H_{macro}$ and $H_{micro}$ of all the known 211, 312, 413 and 514 MAX phase compounds using available published data. Our analysis and comparison show that $Hf_3PB_4$ exhibits the highest hardness values among the MAX phase nanolaminates. Furthermore, we have also calculated the melting temperature of the known MAX phases and found that $Hf_3PB_4$ exhibits the highest melting temperature.

## 2. Computational methodology

The structural, mechanical, electronic, optical and thermal properties of $Hf_3PB_4$ are calculated using density functional theory based on the plane-wave pseudopotential method which is implemented in the CAmbridge Serial Total Energy Package (CASTEP) code [53,54]. The generalized gradient approximation (GGA) of the Perdew–Burke–Ernzerhof (PBE) [55] scheme was used for the exchange and correlations terms. The pseudo-atomic calculations were performed for B - $2s^2$ $2p^1$, P - 3$s$2 3$p$3, and Hf - $5d^2$ $6s^2$ electronic orbitals. A k-point [56] mesh of size 9 × 9 × 2 and a cutoff energy of 500 eV was used. The atomic configuration was optimized with Broyden Fletcher Goldfarb Shanno (BFGS) technique [57] and electronic structure was calculated using electronic density mixing. The self-consistent convergence of the total energy was set to $5 \times 10^{-6}$ eV/atom, and the maximum force on the atom was 0.01 eV/Å. The maximum ionic displacement was set to $5 \times 10^{-4}$ Å, and a maximum stress of 0.02 GPa was used.

## 3. Results and discussion

### *3.1 structural properties*

Fig. 1 shows the unit cell of the $Hf_3PB_4$ belonging to the hexagonal system with space group:$P\bar{6}m2$, (No. 187) [48]. This structure is not like the conventional MAX phases containing C or N but exhibits a number of similar characteristics of conventional MAX phases [48]. The unit cell consists eight atoms in which there are four B atoms, three Hf atoms and only one P atom. The atomic positions of Hf, P and B atoms along with optimized lattice parameters are presented in Table 1. A very high accuracy of our calculation is observed regarding lattice constants. The lattice constants $a$ and $c$ of $Hf_3PB_4$ obtained here are just 0.653 % and 0.583 % higher than the reported values as given in the supporting information of Ref. [48]. Unlike conventional MAX phases, layer of B atoms (2D layer) contribute significantly to the overall stability of the structure. This layer is sandwiched between two Hf layers (Fig. 1) and the boron layer looks like a graphene-layer [48]. The covalent B-B bonding present in the boron layer is much stronger than the M-X bonding as present in the conventional MAX phases, as a consequence the structure of $Hf_3PB_4$ becomes much stiffer compared to many other MAX compounds.

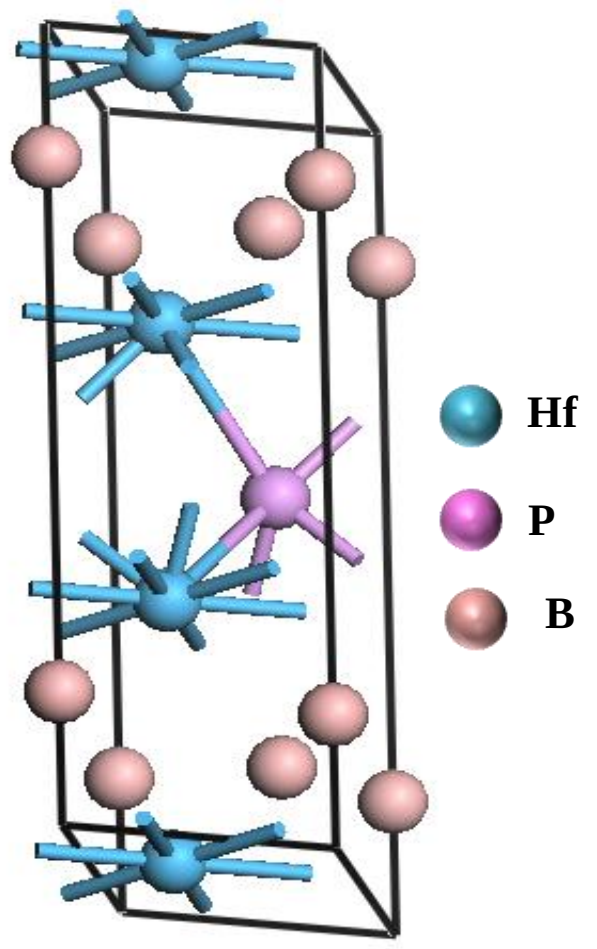

Fig. 1. The schematic unit cell of $Hf_3PB_4$ boride.

**Table 1-** The lattice constants and atomic positions of $Hf_3PB_4$ boride.

| Phases | *a* (Å) | % of deviation | *c* (Å) | % of deviation | | Atomic positions | | |
|---|---|---|---|---|---|---|---|---|
| | | | | | | *x* | *y* | *z* |
| $Hf_3PB_4$ | 3.249085 | 0.653 | 10.520033 | 0.583 | Hf | 0.333333 | 0.666667 | 0.000000 |
| | 3.228[a] | | 10.459[a] | | Hf | 0.333333 | 0.666667 | 0.327468 |
| | | | | | P | 0.666667 | 0.333333 | 0.500000 |
| | | | | | B | 0.000000 | 0.000000 | 0.166419 |
| | | | | | B | 0.666667 | 0.333333 | 0.163670 |

[a]Ref.[48]

### *3.2 Mechanical properties*

The mechanical properties of the predicted $Hf_3PB_4$ are further investigated by calculating the elastic constants through the strain–stress method, which are adapted in the CASTEP module. The methodology has been described in some previous reports [58–62]. The obtained results are presented here under different sub-sections.

*The stiffness constants: mechanical stability*

The mechanical properties of a material are evaluated by the stiffness constants ($C_{ij}$). The $C_{ij}$ are also related to the bonding characteristics, providing with knowledge regarding bonding strength along different crystallographic directions. The stiffness constants also help us to validate the mechanical stability, obtain polycrystalline elastic moduli, understand ductile/brittle nature, machinability index, anisotropic behavior, hardness etc. The compound under study belongs to the hexagonal system; hence they have six stiffness constants: $C_{11}$, $C_{12}$, $C_{13}$, $C_{33}$, $C_{44}$ and $C_{66}$. The last one is not independent and can be calculated using the relation: $C_{66} = (C_{11}-C_{12})/2$. For mechanical stability, the crystal must satisfy the following relations: $C_{11} > 0$, $C_{33} > 0$, $C_{11}-C_{12} > 0$, $C_{44} > 0$, $(C_{11} + C_{12})C_{33} - 2(C_{13})^2 > 0$[63]. As seen in Table 2, the above conditions are satisfied by the $Hf_3PB_4$, therefore, the studied boride is mechanically stable. Since, this is the first calculation of the mechanical properties; therefore, comparison is not possible at this time. Due to this lack of prior information on 314 designated $Hf_3PB_4$, the mechanical properties of some existing Hf based 312 MAX phases are presented in table for comparison.

**Table 2** - The elastic constants, $C_{ij}$ (GPa), bulk modulus, $B$ (GPa), shear modulus, $G$ (GPa), Young's modulus, $Y$ (GPa), machinability index, $B/C_{44}$, macro hardness, $H_{macro}$ (GPa), micro hardness, $H_{micro}$ (GPa), Pugh ratio, $G/B$, Poisson ratio, ν and Cauchy Pressure, $CP$ (GPa) of $Hf_3PB_4$.

| Phase | $C_{11}$ | $C_{12}$ | $C_{13}$ | $C_{33}$ | $C_{44}$ | $B$ | $G$ | $Y$ | $B/C_{44}$ | $H_{macro}$ | $H_{micro}$ | $G/B$ | $v$ | Cauchy Pressure |
|---|---|---|---|---|---|---|---|---|---|---|---|---|---|---|
| $Hf_3PB4$ | 433 | 89 | 140 | 419 | 219 | 225 | 180 | 426 | 1.03 | 29.14 | 37.89 | 0.80 | 0.18 | -130 |
| [a]$Hf_3AlC_2$ | 347 | 77 | 80 | 291 | 127 | 162 | 127 | 302 | 1.28 | 22.59 | 26.31 | 0.78 | 0.19 | -50 |
| [b]$Hf_3SiC_2$ | 348 | 101 | 120 | 335 | 144 | 190 | 127 | 312 | 1.32 | 18.24 | 23.14 | 0.67 | 0.23 | -43 |
| [b]$Hf_3SnC_2$ | 326 | 96 | 97 | 300 | 107 | 170 | 110 | 272 | 1.59 | 15.80 | 19.52 | 0.65 | 0.23 | -11 |

[a]Ref.[64], [b]Ref.[65]

The elastic tensors $C_{11}$ and $C_{33}$ provide with information about the stiffness (elastic) of a material for applied stresses along the <100> and <001> directions, respectively. Therefore, the $Hf_3PB_4$ is the stiffest among the compounds presented in Table 2. Another important stiffness constant, $C_{44}$ providing information regarding hardness directly [66], is also much higher than that of the other compounds presented in Table 2, indicating that the $Hf_3PB_4$ is the hardest material compared to others. Moreover, Aryal et al. [65] have calculated the elastic constants of large numbers known and hypothetical MAX phases including those belonging in the 211, 312, 413 and 514 classes and none have a $C_{44}$ higher than the value of 219 GPa for $Hf_3PB_4$. Based on the values of $C_{44}$ it can be predicted that the $Hf_3PB_4$ is the hardest MAX phase known so far (either synthesized or predicted till now). It should stressed that a slightly larger value of $C_{44}$ (220 GPa) for $Ta_2GeC$ [67] (see Table 1; supplementary information) was reported where LDA was used within DFT. LDA has the tendency of overestimating the elastic constants. We have checked the calculations presented in Ref. [67] with GGA keeping the cut-off energy and k-points grid the same and found $C_{44}$ =178 GPa for $Ta_2GeC$. Moreover, the value of $C_{44}$ turns out to be 244 GPa for $Hf_3PB_4$ when LDA is incorporated in the DFT. In fact, the $C_{11}$, $C_{33}$ and $C_{44}$ values of $Hf_3PB_4$ are much higher than those of the 312 and 413 MAX phases [68–70]. In addition, the unequal values of $C_{11}$ and $C_{33}$ ($C_{11}$ is slightly larger than $C_{33}$) confirms that the bonding strength along *a* and *c* axes are different.

*Elastic moduli: machinability index and hardness*

The stiffness constants are further used to calculate the polycrystalline elastic moduli (*B*, *G*, and *Y*), machinability index ($B/C_{44}$) and macro ($H_{macro}$) and micro ($H_{micro}$) hardness parameters of $Hf_3PB_4$. The bulk modulus (*B*), shear modulus (*G*), Young's modulus (*Y*) are calculated using the well known formalisms [71–74]. As seen from Table 2 that the values of *B*, *G*, and *Y* are also higher than those of the Hf based 312 MAX phases presented here, as a consequence, the $Hf_3PB_4$ has highest resistance to compression and plastic deformation as well as being much stiffer compared to other Hf-based 312 MAX phases [68].

The ratio $B/C_{44}$ [75] measures an important performance indicator known as machinability index. The $B/C_{44}$ ratio is 1.03, lower than other 312 MAX phases. The machinability of a solid is related to its hardness, it is easier to give different shape to a softer solid. In this case, low value of $B/C_{44}$ ratio is a consequence of the high value of $C_{44}$. For example, the highest machinability (33.33) is found for $W_2SnC$ with the lowest $C_{44}$ (6 GPa) among all the 211 MAX phases [76]. On the other hand, the highest $B/C_{44}$ ratio is found for $Ti_3InC_2$ with the lowest $C_{44}$ (92 GPa) [77] among the 312 MAX phases.

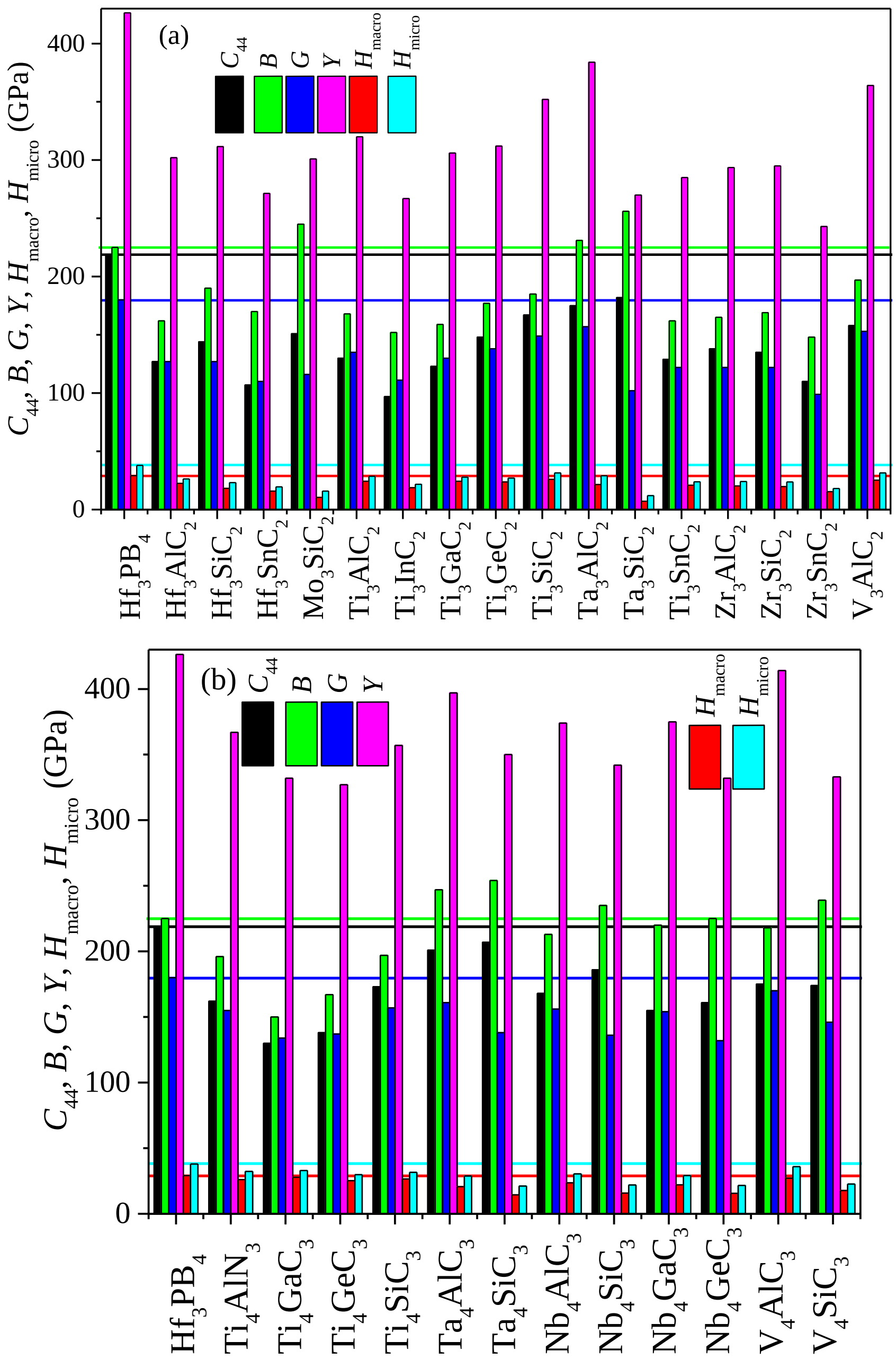

Fig. 2. Comparison of $C_{44}$, $B$, $G$, $Y$, $H_{\text{macro}}$ and $H_{\text{micro}}$ of $Hf_3PB_4$ boride with those of (a) 312 and (b) 413 known MAX phases.

We have made use of the computed elastic moduli to compare the hardnesses of different MAX compounds using the following relations: $H_{\text{macro}} = 2\left[\left(\frac{G}{B}\right)^2 G\right]^{0.585} - 3$ [78] and $H_{\text{micro}} = \frac{(1-2\nu)E}{6(1+\nu)}$[74]. These hardness parameters are useful tools to compare the hardness among the materials belonging to the same family. The listed values of ($H_{\text{macro}}$) and micro ($H_{\text{micro}}$) of $Hf_3PB_4$ are higher than those of known MAX phases. Fig. 2 shows a comparative picture of $H_{\text{macro}}$ and $H_{\text{micro}}$ for $Hf_3PB_4$ along with the $H_{\text{macro}}$ and $H_{\text{micro}}$ of (a) 312 and (b) 413

MAX phases. As can be seen in the figure both $H_{macro}$ and $H_{micro}$ for $Hf_3PB_4$ are higher than those 312 and 413 MAX phases. We have also listed the published elastic properties of ninety one (91) 211 MAX phase carbides, nitrides and borides. Using the published data, we have calculated the $H_{macro}$ and $H_{micro}$ for those ninety one (91) 211 MAX phases given in the Table 1 of supplementary information and found that none have higher $H_{macro}$ and/or $H_{micro}$ than that of $Hf_3PB_4$. We have found the elastic properties of only two 514 MAX phases, $Ti_5AlC_4$ and $Ti_5SiC_4$ so far with lower values of $H_{macro}$ (25.31 and 27.11 GPa, respectively) and $H_{micro}$ (29.84 and 34.19 GPa, respectively). Another point should be mentioned here that the bulk moduli (*B*) of some 312 and 413 MAX phases as seen in Fig. 2, have larger values than that of $Hf_3PB_4$. The bulk moduli of some 211 MAX phases have also larger values than that of $Hf_3PB_4$. Although, the values of $C_{44}$, *B*, *G*, and *Y* provide information regarding the mechanical strength of solids but only $C_{44}$ and *G* are closely related with the hardness of the material compared to *B* and *Y*. Jhi et al. [66] have explained a close relationship between hardness and $C_{44}$ for binary carbides and nitrides. The equations used to calculate the $H_{macro}$ and $H_{micro}$ also provide the information regarding relationship among *B*, *G*, *Y*, and hardness. Thus, it is clear that the high *B* value does not warrant high hardness for solids. It is noteworthy that the *Y* of $Hf_3PB_4$ is also large compared to 312 and 413 MAX phases (Fig. 2) as well as 211 MAX phases (Table 1; supplementary information).

The Vickers hardness attributed from the hardness of all the bonds existing within a solid is also calculated for $Hf_3PB_4$ by employing Mulliken bond population analysis. The Vickers hardness can be calculated using Gao method [77] which is applicable for non-metallic solids dominated by covalent bond. This method has been extended for partially metallic bonded compounds by further work by Gou et al. [79]. This modified scheme has been used successfully to estimate Vickers hardness of a large number of compounds with metallic character including MAX and MAB phase nanolaminates [9,11,52,58,80]. The relevant formula for the hardness is given as [80]: $H_V = \left[\prod^{\mu}\left\{740\left(P^{\mu} - P^{\mu'}\right)\left(v_b^{\mu}\right)^{-5/3}\right\}^{n^{\mu}}\right]^{1/\sum n^{\mu}}$, where $P^{\mu}$ is the Mulliken population of the $\mu$-type bond, $P^{\mu'} = n_{free}/V$ is the metallic population, and $v_b^{\mu}$ is the bond volume of $\mu$-type bond. The obtained Vickers hardness of $Hf_3PB_4$ is 7.85 GPa (Table 3) which is higher than most of the conventional MAX phases (4.9 GPa of $Hf_3AlC_2$[64], 4.7 GPa of $Hf_3SnC_2$ [81]). Although, some authors have reported higher values of Vickers hardness for some 211 MAX phases ($Mo_2GaC$ - 9.6 GPa [82]; $Nb_2InC$ - 7.9 GPa [82]; $Hf_2PN$ - 11.5 GPa [83]) but the their elastic constant and moduli ($C_{44}$, $B$, $G$, and $Y$) related to the hardness are still lower than those of $Hf_3PB_4$. The Vickers hardness ($H_v$) for MAX phases generally resides within the limit 2-8 GPa [64]. Though, the Vickers hardness of $Hf_3PB_4$ is high compared to most of the existing MAX phases [9,11,64,81,84] but it is still not so high, indicating a machinable characteristics of $Hf_3PB_4$ like many other MAX phase compounds. The Vickers hardness of MAX compounds is seldom reported. Therefore, a complete comparison of the Vickers hardness is not possible at this stage. The hardness value obtained by experiments usually depends on the measurement techniques. The theoretically obtained values of hardness also depend on the formalism used for its calculation and we have used different formalisms to calculate $H_{macro}$, $H_{micro}$ and $H_v$, consequently, different values are obtained. However, based on the values of $C_{44}$, $H_{macro}$, and $H_{micro}$ one can infer that $Hf_3PB_4$ is the hardest among the MAX phases known so far.

In summary, the important mechanical parameters related to materials hardness of $Hf_3PB_4$ obtained so far are higher than those of the known conventional MAX phases. Now, the question that naturally arises is what is the physical process behind these high values of mechanical parameters of $Hf_3PB_4$? To get the answer, we look closely at the chemical composition of the compound under study. For 314 phases (like $Hf_3PB_4$), the X element (B) is two times the X element (C/N) present in the conventional 312 MAX phases. This additional B atom results in the strong B–B covalent bonds [49] as seen in Table 3; consequently, a stiffer structure of 314 phases compared to conventional 312 MAX phases emerges. This statement will be further strengthened in Section 3.3 with the aid of charge density mapping.

Table 3 - Calculated Mulliken bond, bond length $d^{\mu}$, bond overlap population $P^{\mu}$, metallic population $P^{\mu'}$, bond volume $v_b^{\mu}$, bond hardness $H_v^{\mu}$ of $\mu$-type bond and Vickers hardness, $H_v$ of $Hf_3PB_4$.

| Bond | $d^{\mu}$ (Å) | $P^{\mu}$ | $P^{\mu'}$ | $v_b^{\mu}$ (Å$^3$) | $H_v^{\mu}$ (GPa) | $H_v$(GPa) |
|---|---|---|---|---|---|---|
| B-B | 1.87608 | 2.50 | 0.0338 | 3.5100 | 225.1 | 7.85 |
| B-Hf | 2.52770 | 0.29 | 0.0338 | 9.2607 | 4.640 | |
| B-Hf | 2.5462 | 0.14 | 0.0338 | 8.77539 | 2.139 | |
| B-Hf | 2.5471 | 0.18 | 0.0338 | 8.78480 | 2.927 | |
| B-Hf | 2.5659 | 0.12 | 0.0338 | 8.98002 | 1.677 | |
| P-Hf | 2.61022 | 1.10 | 0.0338 | 48.088 | 1.240 | |

*The brittleness of $Hf_3PB_4$*

The Pugh ratio (*G*/*B*), Poisson's ratio ($\upsilon$) and Cauchy pressure (*CP*), ($C_{12}$-$C_{44}$), are well known tools used to separate the solids into brittle or ductile groups where the critical value of *G*/*B* ratio is 0.571 [85], $\upsilon$ is 0.26 [64] and *CP* is zero [86]. The values of *G/B* greater than 0.571, υ lower than 0.26 and negative values of *CP* signify the brittleness of solids. The listed values (*G/B* = 0.80, υ = 0.18 and *CP* = -130) signify that the $Hf_3PB_4$ is brittle a material. Beside, the magnitude of negative *CP* also provides information regarding the dominant role of covalent bond within the solids. The magnitude of *CP* of $Hf_3PB_4$ is much higher than that of 312 MAX phases considered here, suggesting the presence of much stronger covalent bonds within $Hf_3PB_4$ in comparison to other MAX compounds.

### ***3.3 The elastic anisotropy***

Study of mechanical anisotropy carries significant interest owing to its close relation with a number of important physical processes [87]. Motivated by these points [86,87], the anisotropy of $Hf_3PB_4$ has been studied by means of different anisotropy indices. First of all, the direction dependencies of Young's modulus, compressibility, shear modulus and Poisson's ratio are studied using the ELATE code [88] by plotting the values in 2D and 3D presentations.

The level of the anisotropy can be understood from the nature of the profiles of the 3D and 2D planar projection plots of the elastic moduli and constants. For isotropic solids, completely spherical (3D) and circular (2D) profiles are expected; while the degree of distortion from spherical and circular shapes implies the degree of anisotropy in elastic moduli and constants of solids. Fig. 3(a) demonstrates the direction dependence of Young's modulus (*Y*). As seen from Fig. 3(a), a circular presentation of *Y* in the *xy* plane confirms its isotropic nature whereas anisotropic nature is observed in the *xz* and *yz* planes. *Y* is minimum (343.66 GPa) on both *xz* and *yz* planes and is maximum at 45° (470.45 GPa) of the axes on both *xz* and *yz* planes; the anisotropic ratio is, therefore, 1.37.

Table 4 - The minimum and the maximum values of the Young's modulus, compressibility, shear modulus, and Poisson's ratio of $Hf_3PB_4$.

| $Y_{min.}$ (GPa) | $Y_{max.}$ (GPa) | $A_Y$ | $K_{min}$ ($TPa^{-1}$) | $K_{max}$ ($TPa^{-1}$) | $A_K$ | $G_{min.}$ (GPa) | $G_{max.}$ (GPa) | $A_G$ | $\upsilon_{min.}$ | $\upsilon_{max.}$ | $A_\upsilon$ |
|---|---|---|---|---|---|---|---|---|---|---|---|
| 343.66 | 470.45 | 1.37 | 1.34 | 1.55 | 1.15 | 140.94 | 218.78 | 1.55 | 0.071 | 0.298 | 4.20 |

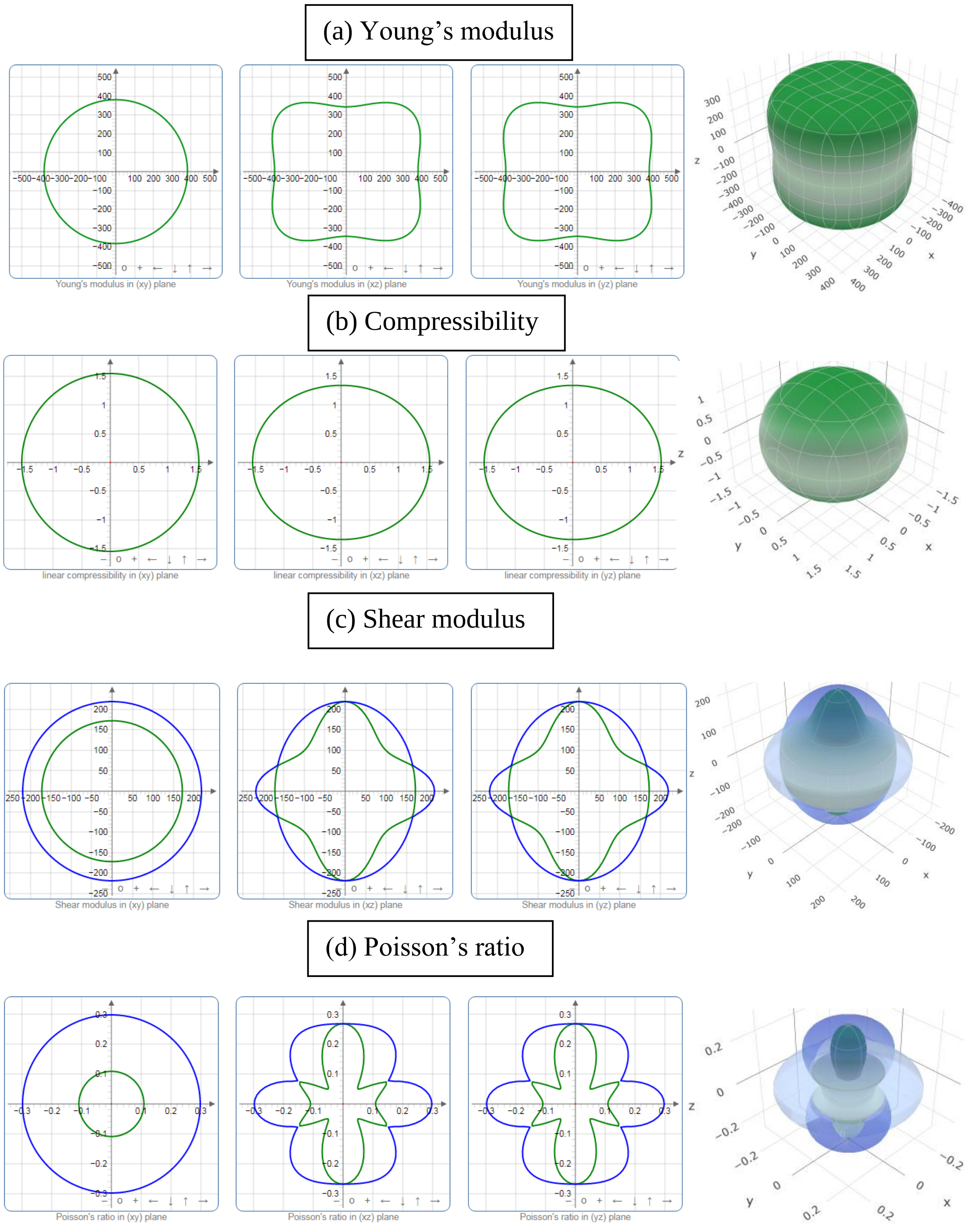

Fig. 3. The 2D and 3D plots of (a) Y, (b) K, (c) G and (d) υ of $Hf_3PB_4$.

Fig. 3(b) shows the anisotropy in compressibility (*K*); a circular presentation of *K* in *xy* plane confirms its isotropic nature whereas slightly anisotropic nature is observed in the *xz* and *yz* planes. Fig. 3(c) illustrates the shear modulus along different directions. As seen from Fig. 3 (c), the maximum value (218.78 GPa) of shear modulus is obtained on both *xz* and *yz* planes and minimum value (140.94 GPa) is observed at around 45° from the *z*-axis with a perfect circle represents the isotropic nature of shear modulus in the *xy* plane. Fig. 3(d) exhibits the direction dependency of Poisson's ratio ($v$). Like the other elastic moduli, isotropy is observed in the *xy* plane but complex and systematic anisotropic nature with a maximum value on the vertical axis is observed in *xz* and *yz* planes. The minimum and maximum values of $Y$, $K$, $G$ and $\upsilon$ of $Hf_3PB_4$ are enlisted in Table 4.

The stiffness constants ($C_{ij}$) can also be used to explore some anisotropic indices. The shear anisotropic factors $A_i$ (i = 1, 2 and 3) can be measured by the following formulae: $A_1 = \frac{1/6(C_{11}+C_{12}+2C_{33}-4C_{13})}{C_{44}}$, $A_2 = \frac{2C_{44}}{C_{11}-C_{12}}$, $A_3 = A_1.A_2 = \frac{1/3(C_{11}+C_{12}+2C_{33}-4C_{13})}{C_{11}-C_{12}}$ [89] for the {100}, {010} and {001} planes in between $\langle 011\rangle$ and $\langle 010\rangle$, $\langle 101\rangle$ and $\langle 001\rangle$, and $\langle 110\rangle$ and $\langle 010\rangle$ directions, respectively. The values of bulk modulus along $a$ and $c$-directions can be obtained using the following relations [90]: $B_a = a\frac{dP}{da} = \frac{\Lambda}{2+\alpha}$,$B_c = c\frac{dP}{dc} = \frac{B_a}{\alpha}$,where $\Lambda = 2(C_{11} + C_{12}) + 4C_{13}\alpha + C_{33}\alpha^2$and $\alpha = \frac{(C_{11}+C_{12})-2C_{13}}{C_{33}+C_{13}}$. Finally, the anisotropy in linear compressibility coefficients ($k_c/k_a$) along the $a$- and $c$-axis can be expressed by the equation [91]: $\frac{k_c}{k_a} = C_{11} + C_{12} - 2C_{13}/(C_{33} - C_{13})$. All these anisotropic factors are calculated and listed in Table 5. The obtained values of $A_i$'s ($A_i$ = 1 for isotropic nature); $B_a$ and $B_c$ ($B_a = B_c$ for isotropic nature) and $k_c/k_a$ ($k_c/k_a$ = 1 for isotropic nature) revealed the overall mechanical/elastic anisotropic behavior of $Hf_3PB_4$.

Table 5 - Anisotropic factors, $A_1$, $A_2$, $A_3$, $k_c/k_a$, $B_a$, $B_c$ and universal anisotropic index $A^U$ of $Hf_3PB_4$.

| $A_1$ | $A_2$ | $A_3$ | $k_c/k_a$ | $B_a$ | $B_c$ | $A^U$ |
|---|---|---|---|---|---|---|
| 0.61 | 1.27 | 0.78 | 0.87 | 643 | 741 | 0.177 |

In Table 5, the universal anisotropic index $A^U$ is shown which was obtained using the upper bound (Voigt, $V$) and lower bound (Reuss, $R$) of $B$ and $G$ following the equation [92] $A^U = 5\frac{G_V}{G_R} + \frac{B_V}{B_R} - 6 \geq 0$. The non-zero value of $A^U$ implies the anisotropic nature of $Hf_3PB_4$ ($A^U$ = 0 for isotropic solids). In summary, one can conclude that like all other MAX phase nanolaminates, $Hf_3PB_4$ is anisotropic in nature.

### *3.4 Electronic properties*

*Electronic band structure and density of states:* We have calculated the electronic band structure and density of states (DOS) of $Hf_3PB_4$ as shown in Fig. 4(a) and (b). Like the conventional MAX phases, the metallic nature is confirmed from the overlapping of valence band and conduction band together with a finite value of DOS (1.42 states/eV) at the Fermi level. Our result is in accord with the results obtained by Miao et al. [48]. The anisotropic nature of electronic conductivity is seen from the difference in the energy dispersion. The energy dispersion along the *c*-direction, e.g., along Γ-A, H-K and M-L are much less compared to the energy dispersion in the basal plane such as along A-H, K-Γ, Γ-M and L-H, as shown in Fig. 4(a). The reason for this anisotropy is the difference in the electronic effective mass tensor which is lower in *ab*-plane compared to that along the *c*-direction [93]. Thus, it can be concluded that the conductivity is lower along the *c*-direction compared to that in the *ab*-plane for $Hf_3PB_4$. This feature is also true for other conventional MAX phases [94]. To get better insights of the bonding characteristics we have calculated total and atomic orbital resolved partial DOS of $Hf_3PB_4$ as shown in Fig. 4(b). The contribution from different electronic states becomes clear from the partial DOS plots. The DOS at the Fermi level is attributed from Hf-*d*, P-*p* and B-*p* states with a dominant role of Hf-*d* states. The contribution from B-*p* states is smaller in magnitude but the contribution from P-*p* states can be ignored. The partial DOS is also used to demonstrate the hybridization between different states. The peaks in the DOS are the consequences of hybridization among the different electronic states observed in different energy ranges. The peak in the low energy region -9 eV to -13 eV is attributed from the hybridization of B-2*s* and P-3*s* states with a major involvement of B-2*s* states. The energy region -2.5 eV to -7.5 eV exhibits the peak that results from the hybridization of B-2*p* and Hf-5*d* states. Just below the Fermi level, the hybridization of Hf-5*d* and P/B-*p* states results in the peak. The bonds results from the hybridization among different states are mainly covalent bonding. The conduction band is dominating by the metal-to-metal *dd* interactions and anti-bonding states. Qualitatively similar electronic band structure features are also observed for conventional MAX phases.

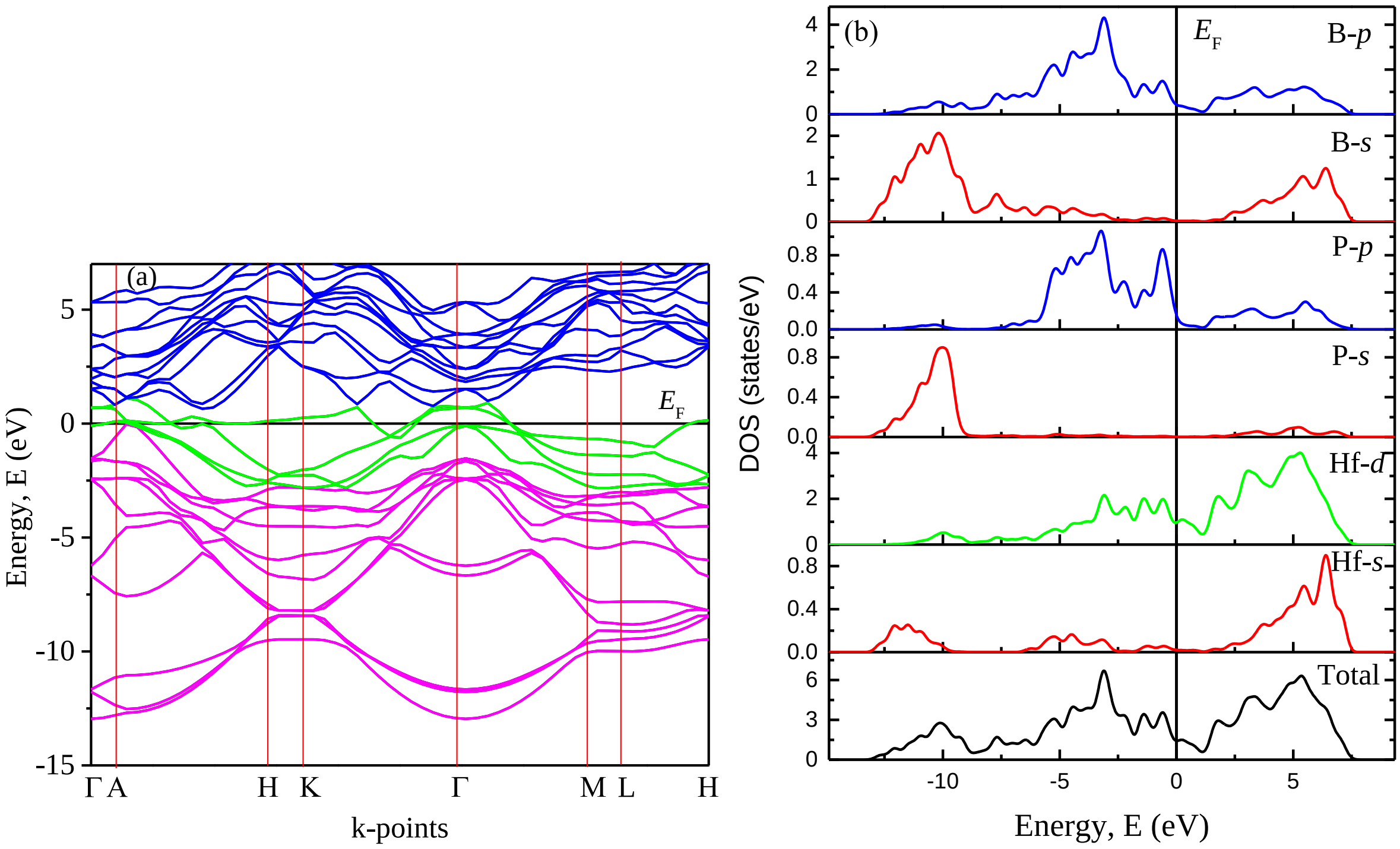

Fig. 4. (a) Electronic band structure and (b) density of states (DOS) of $Hf_3PB_4$.

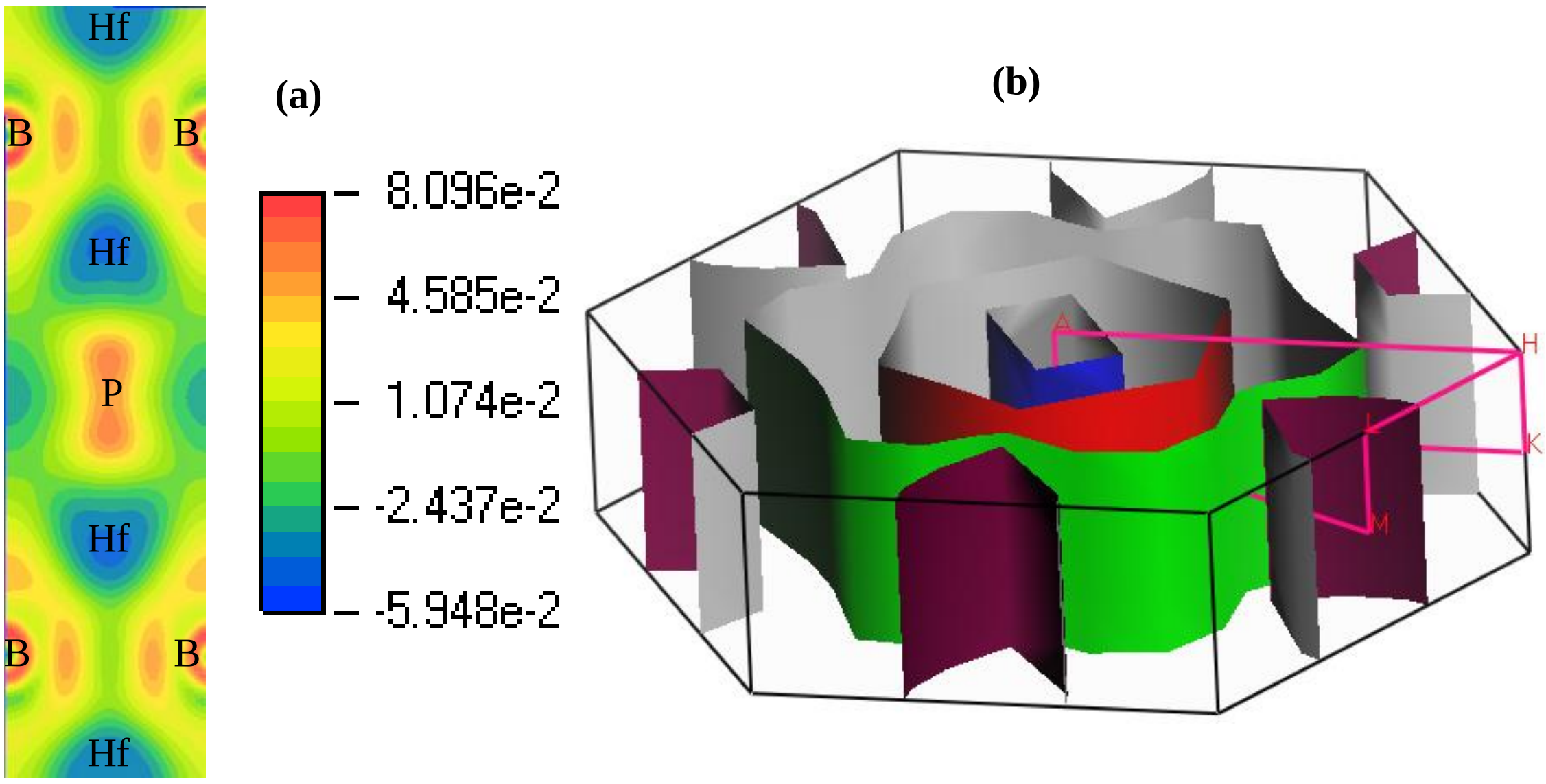

Fig. 5. (a) Charge density mapping image and (b) Fermi surface of $Hf_3PB_4$ compound.

*Charge density mapping (CDM):* The charge density mapping exposes the electron densities involved in the chemical bonds between atoms. The CDM exhibits two regions: the accumulated region of charges and depleted region of charges. The covalent bonding is attributed by the bonding of the accumulated charges in between two atoms, whereas the balancing of the accumulated and depleted regions results in ionic type bonding. As seen from the figure (Fig. 5(a)) that the charges are accumulated between the adjacent boron atoms, therefore, existence of strong covalent bonding B-B (two center-two electron (2c-2e)) is expected within $Hf_3PB_4$. This type of bonding is also reported for $Ti_2InB_2$ [49]. For $Ti_2InB_2$, Wang et al. explained that the charge (0.87 $|e|$) is transferred from Ti atom to B atom, leads the formation of 2c-2e bond between B atoms. The Mulliken analysis revealed that 0.74 $|e|$ charge is transferred from Hf atom to B atom, therefore, 2c-2e bond between B atoms is also expected for $Hf_3PB_4$ boride. The formation of 2c-2e bond can be understood as follows. As confirmed from the Mulliken analysis that 0.74$|e|$ charge is transferred from Hf to B atoms. One can see (Fig. 1) that within the B layers, there is a B atom at the interstitial position and there are four B atoms at the edges of the unit cell. The transferred charge is assumed to be shared between the interstitial B atom and edged B atoms, consequently, 2c-2e bond is formed between the interstitial B atoms and edged B atoms. The bonding strength can also be realized from the Vickers hardness of individual bonds as presented in Table 3. Another comparatively weaker bond is expected between the charges accumulated at positions of Hf and P atoms. The weakest covalent bond is observed between Hf and B atoms.

*Fermi surface:* Fig. 5 (b) shows the Fermi surface of $Hf_3PB_4$ in which both electron- and hole-like sheets are present. The four sheets are the consequences of the four bands crossing the Fermi level. The inner three sheets are cylindrically co-axial having different cross-sectional views. The 1$^{st}$ sheet has square cross-section while 2$^{nd}$ sheet has hexagonal cross-section. The 3$^{rd}$ sheet has complicated hexagonal cross-section which is expanded along *G-K* direction and shrunk along *G-M* direction of the Brillouin zone. The 4$^{th}$ sheet is composed of six up curved type ribbon centered at *M-L* direction of the Brillouin zone. The Fermi surface of $Hf_3PB_4$ is due to low dispersive Hf-5d states.

In summary, the bonding nature explained by DOS and CDM disclosed why the mechanical properties of $Hf_3PB_4$ are much enhanced than those of other MAX phases.

### *3.5 Thermal properties*

The important thermodynamic parameters such as Debye temperature, minimum thermal conductivity, Grüneisen parameter and melting temperature of $Hf_3PB_4$ are studied in this section.

The Debye temperature ($\Theta_D$) is one of the fundamental characteristic parameter of solids that assists to understand a number of important physical parameters through the correlation between mechanical properties and thermodynamic properties interlinking lattice dynamics. For example phonons, lattice vibration enthalpy, thermal conductivity, melting point, specific heat etc [93,95]. Anderson [96] developed a simple method to calculate $\Theta_D$ using average sound velocity ($v_m$) which is correlated with shear modulus ($G$) and bulk modulus ($B$). The $v_m$ can be estimated by the following expression:

$v_m = \left[1/3\left(1/v_l^3 + 2/v_t^3\right)\right]^{-1/3}$ where, $v_l$ and $v_t$ be longitudinal and transverse sound velocity, respectively. Again, $v_l$ and $v_t$ are related to the elastic moduli (shear and bulk modulus) and density of the solid as expressed by the following equations:

$v_l = [(3B + 4G)/3\rho]^{1/2}$ and $v_t = [G/\rho]^{1/2}$.

Finally, $\Theta_D$ can be calculated by the equation: $\Theta_D = {}^{h}/_{k_B}\left[\left({}^{3n}/_{4\pi}\right)N_A\rho/M\right]^{1/3} v_m$, where, $M$ be the molar mass, $n$ be the number of atoms in the molecules, ρ be the mass density, and $h$, $k_B$, and $N_A$ be the Planck's constant, Boltzmann constant and Avogadro's number respectively. The estimated values of crystal density, longitudinal, transverse and average sound velocities ($v_l$, $v_t$, and $v_m$, respectively) and Debye temperature ($\Theta_D$) are presented in Table 6. Like mechanical properties, the obtained $\Theta_D$ of $Hf_3PB_4$ (592 K) is also much higher than that of other Hf-based 312 MAX phases (e.g., for $Hf_3AlC_2$, $\Theta_D$ ~ 459 K [64]; for $Hf_3SnC_2$, $\Theta_D$ ~405 K [81]). Such high value of $\Theta_D$ is entirely consistent with the remarkable hardness related parameters obtained for $Hf_3PB_4$ in preceding sections. Quite generally $\Theta_D$ is higher for harder solids and vice versa [72,93]. Though $\Theta_D$ of $Hf_3PB_4$ (592 K) is lower than some of the other MAX phase nanolaminates but it is higher than that of other Hf-based MAX phases (see Table 5; supplementary information); it is due to the heavy molecular mass rather than due to the stiffness of the crystal. In particular, the lower value of $\Theta_D$ might be due to the presence of Hf. For instance, Bouhemadou et al. [97] have calculated the Debye temperature of $Ti_2SC$ (800 K), $Zr_2SC$ (603 K) and $Hf_2SC$ (463) even though their elastic moduli are in comparable order (see Table 1; supplementary information). Another important factor can be noticed from Table 6 that the values of $v_l$ is much higher than than that of $v_t$ [98]; the wave velocity in transvers mode is reduced owing to the loss of more energy to vibrate neighboring atoms compared to that during the propagation of longitudinal mode [99].

**Table 6**

Calculated density ($\rho$), longitudinal, transverse and average sound velocities ($v_l$, $v_t$, and $v_m$, respectively), Debye temperature ($\Theta_D$), minimum thermal conductivity ($K_{min}$), Grüneisen parameter ($\gamma$) and melting temperature ($T_m$) of $Hf_3PB_4$.

| $\rho$ (g/cm$^3$) | $v_l$ (m/s) | $v_t$ (m/s) | $v_m$ (m/s) | $\Theta_D$ (K) | $K_{min}$ (W/mK) | $\gamma$ | $T_m$ (K) |
|---|---|---|---|---|---|---|---|
| 10.52 | 6646 | 4135 | 4557 | 592 | 1.14 | 1.23 | 2282 |

Due to their promising properties, MAX phases are also considered as potential materials for high temperature applications. Therefore, the technologically important parameters, such as, $K_{min}$, $\gamma$ and $T_m$ of $Hf_3PB_4$ are calculated because of their engineering importance. The minimum thermal conductivity ($K_{min}$) defines a constant thermal conductivity at higher temperature when the phonon contribution to the thermal conductivity ($K_{ph}$) reaches its minimum value and becomes independent of temperature. $K_{min}$ has been estimated using the equation [100]: $K_{min} = k_B v_m \left(\frac{M}{n\rho N_A}\right)^{-2/3}$, where $k_B$, $v_m$, $N_A$ and $\rho$ are Boltzmann constant, average phonon velocity, Avogadro's number and density of crystal, respectively, and is given in Table 6. The value of $K_{min}$ is comparable with the $K_{min}$ of other MAX phases [7,23]; the low value of $K_{min}$ and high melting point suggest the possible use as thermal barrier coating (TBC) material for high temperature applications although value of thermal expansion coefficient remains still unexplored.

The Grüneisen parameter ($\gamma$) is an important parameter that provides the knowledge of anharmonic effect in the crystal closely related to the lattice dynamics and has been calculated using the equation based on the Poisson's ratio [101]: $\gamma = \frac{3}{2}\frac{(1+\nu)}{(2-3\nu)}$. For polycrystalline materials, $\gamma$ should lie within the range of 0.85 to 3.53 where the Poisson's ratio should have the value within the range of 0.05–0.46 [102]. The calculated of value of $\gamma$ is 1.23 which is within the established range. Moreover, the obtained value is comparatively low that implies low anharmonic effects in $Hf_3PB_4$. This is directly related to the high stiffness of the compound under study since soft phonon modes contribute significantly in the anharmonicity of solids. The information of melting temperature ($T_m$) is mandatory for materials which are applicable at high temperature. The melting temperature of $Hf_3PB_4$ is calculated from the stiffness constants using the equation [103]: $T_m = 354 + \frac{4.5(2C_{11}+C_{33})}{3}$ and is given in Table 6. The value of $T_m$ (2282 K) is also very high compared to typical Hf-based MAX phase compounds (e.g., for $Hf_3SnC_2$, $T_m$ ~1773 K) [81]. Actually, the melting temperatures for known MAX phases have also been calculated using the published data (see Table 1-4; supplementary information) and we see that none of the other MAX phases have higher $T_m$ than that of $Hf_3PB_4$. Therefore, $Hf_3PB_4$ is highly suitable for high temperature applications.

In summary, like the mechanical properties, the thermal properties of $Hf_3PB_4$ exhibits higher values of thermal parameters and lower level of anharmonicity compared to other MAX phases, suggesting that $Hf_3PB_4$ is better candidate for high temperature applications compared to most of the existing MAX phase nanolaminates.

### *3.6 Optical properties*

In addition to the properties presented in the preceding sections, we have also studied the optical properties of $Hf_3PB_4$ to check the possibility for its use in optical devices as well as coating material. Here we have studied different optical functions such as dielectric constant, refractive index, absorption coefficient, photo-conductivity, reflectivity and loss function in the photon energy range up to 20 eV with two polarization directions of [100] and [001] for the first time using the CASTEP code. The details methodology of optical calculations can be found elsewhere [60,104]. An energy smearing of 0.5 eV has been applied for Gaussian broadening for the calculations of all the optical parameters. As mentioned in Section 3.4, the material is metallic in nature, therefore an unscreened plasma frequency of 3 eV and damping of 0.05 eV were used to compute the low energy part of the dielectric function. It is observed from the all optical parameters that electric field polarization direction has little effect on the optical constant spectra. Therefore, we will discuss the behavior of the optical parameters with respect to the [001] polarization only throughout this section.

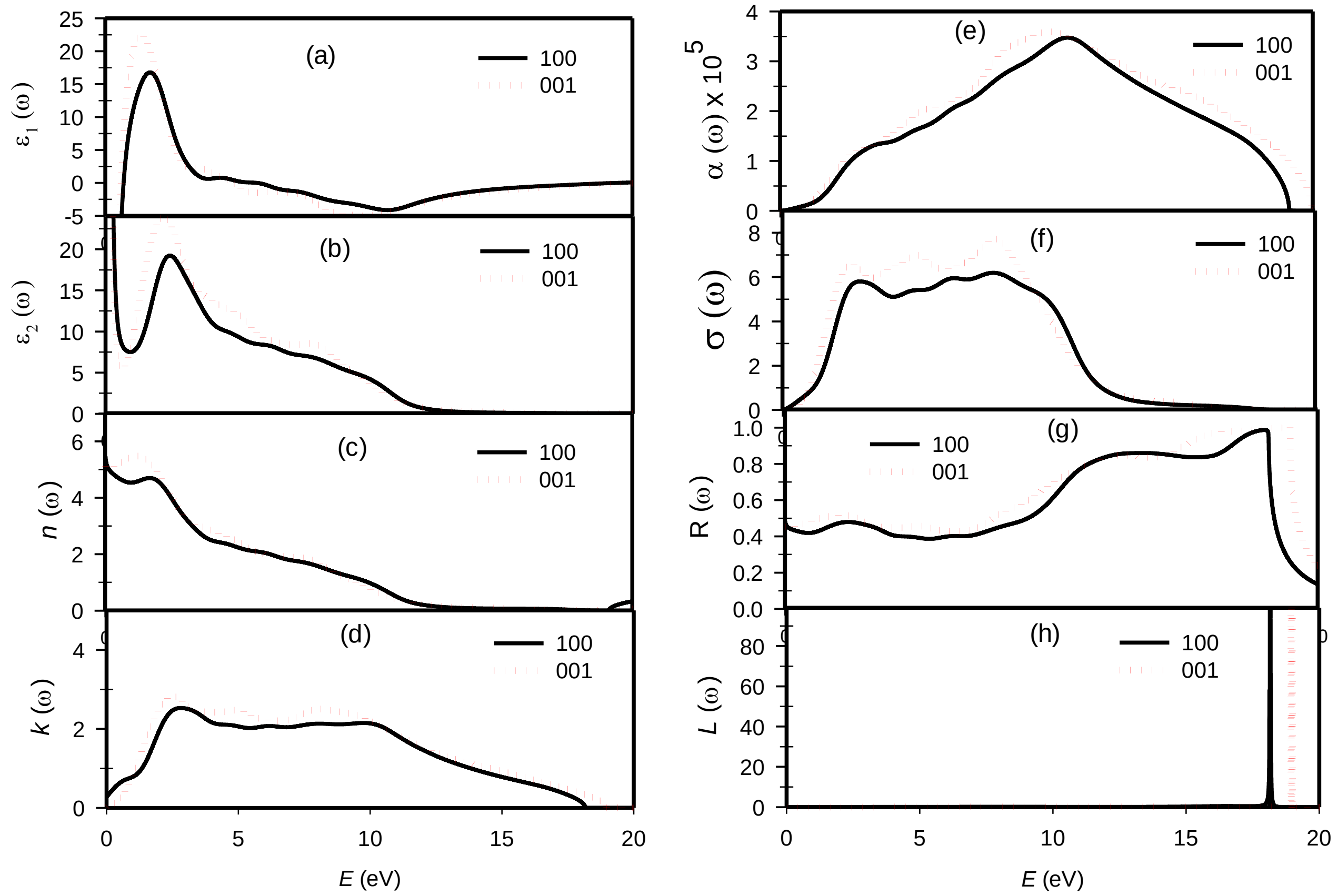

Fig. 6. Different optical functions (a) real part of dielectric function, $\mathcal{E}_1$, (b) imaginary part of dielectric function, $\mathcal{E}_2$, (c) refractive index, *n*, (d) extinction coefficient, *k*, (e) absorption coefficient, *α*, (f) photoconductivity, *σ*, (g) reflectivity, *R* and (h) loss function, *L* as a function of electromagnetic radiation energy.

The real part of dielectric function, $\mathcal{E}_1$ is illustrated in Fig. 6 (a). The spectra of $\mathcal{E}_1$ start from negative value to positive, which is common for metallic systems. The large negative value of $\mathcal{E}_1$ is an indication of supporting the well known Drude model for metallic materials, which is consistent with our previous discussion in Section 3.4. The peak for $\mathcal{E}_1$ at low energy (~ 1 eV) is found due to the intra-band transition of electrons in this compound. The imaginary part of dielectric function $\mathcal{E}_2$ is depicted in Fig. 6 (b). A sharp peak is found at 2 eV and then $\mathcal{E}_2$ gradually decreases and finally goes to zero at 11.60 eV of photon energy. The refractive index (*n*) and extinction coefficient (*k*), which are useful parameters to design various optical devices, are depicted in Fig. 6 (c) and (d), respectively. The static value of *n*(0) was found to be 5.47 which remained almost constant in the long wavelength infrared region and then gradually decreased. On the other hand, the value of *k* gradually increased, reaching a high value of 2.8 at 2.6 eV and then decreased very slowly as shown in Fig. 6 (d). Fig. 6 (e) reveals the variation of absorption spectra as a function of photon energy. It is known that optical absorption is one of the important properties for photovoltaic and optoelectronics devices and it gives knowledge for maximum power conversion efficiency. The highest absorption peak was attained at photon energy of 9.75 eV. The absorption spectra indicate clearly that the studied compound can absorb significant part of solar spectrum (from visible to ultraviolet radiation) quite efficiently. The photoconductivity is another important characterizing parameter for optoelectronic devices. Optical conductivity profile as a function of photon energy is depicted in Fig. 6 (f). The features of optical conductivity curve are almost identical with that of photon absorption profile. It is noted here that the photoconductivity is almost constant when the incident photon energy lies within 2.4 -7.6 eV. The reflectivity ($R$) spectrum as a function of incident photon energy is depicted in Fig. 6 (g). The reflectivity spectra starts with a value of 0.445 (44.5 %) of this compound and this value remained almost constant up to 6.88 eV. This implies that the studied material is capable to reflect solar radiation in the infrared, visible and near-ultraviolet regions. It is reported that if a material has ability to reflect ~ 44 % of visible light, the compound would be capable of reducing solar heating [105]. Therefore, the $Hf_3PB_4$ compound can be used as a coating material for reducing solar heating. The computed energy loss spectra can be seen in Fig. 6 (h). The energy loss spectra of materials are very crucial to understand the screened excitation spectra produced by swift charges inside the material. The loss function can be related to the plasma frequency when a fast electron travels through a material. Indeed, there is no loss of energy when the incident energy is less than 19 eV. The appearance of highest peak of loss spectrum is due to bulk plasmonic excitation at an energy of 19 eV. The frequency corresponding to this particular energy is called the plasma frequency. For energies of photons exceeding the plasma energy, both reflectivity and absorption coefficient decreases sharply approaching zero and the compound becomes transparent to the incident photons.

## 4. Conclusions

The physical properties of recently predicted thermodynamically stable $Hf_3PB_4$ boride have been explored in detail using first-principles calculations. The obtained lattice constants exhibit a very good agreement with the reported values. The Born conditions on stiffness constants confirm the mechanical stability of $Hf_3PB_4$. The $C_{11}$, $C_{33}$ and $C_{44}$, $B$, $G$, $Y$, $H_{macro}$, and $H_{micro}$ of $Hf_3PB_4$ are much larger than those of the Hf-based 312 MAX phases. Some of the mechanical parameters concerning the strength of the material, $C_{44}$, $H_{macro}$, and $H_{micro}$ are compared with known 211, 312, 413 and 514 MAX phases studied till date. We have found that the $Hf_3PB_4$ exhibits the highest values, resulting in the hardest possible MAX compound. The Vickers hardness is also much higher (7.85 GPa) than that of Hf-based 312 MAX phases. We predict that the strong B-B bonding leads such high values of hardness. The brittleness of $Hf_3PB_4$ is confirmed from the Pugh ratio, Poisson's ratio and Cauchy pressure. The direction dependent elastic moduli and anisotropy factors confirm anisotropic nature of $Hf_3PB_4$. The electronic band structure confirms the metallic nature of $Hf_3PB_4$. Analysis of DOS reveals hybridization between electronic orbitals of unlike atoms while the CDM discloses the presence of strong B-B covalent bonding. Both electron and hole like sheets are present in the Fermi surface. The computed Debye temperature (592 K) is also much higher than those of existing Hf-based MAX phases. A low anharmonic effects in $Hf_3PB_4$ is confirmed from the low value of $\gamma$ (1.23). The melting temperature (2282 K) of $Hf_3PB_4$ is higher than that of all the other known MAX phases. The low value of $K_{\min}$ and high melting point suggest the possible use of $Hf_3PB_4$ as a thermal barrier coating (TBC) material in the high temperature device applications. The optical conductivity and absorption coefficient agree with the electronic band structure results. The $Hf_3PB_4$ has potential to be used as a coating material for reducing solar heating. The obtained results suggest that the predicted $Hf_3PB_4$ is a better candidate for applications in many sectors compared to other existing and predicted Hf-based MAX phases. The obtained results of $C_{44,}$ $H_{\mathrm{macro}}$, and $H_{\mathrm{micro}}$ of $Hf_3PB_4$ encourage us to predict that the $Hf_3PB_4$ boride should be the hardest among the MAX phases known so far. We hope that the present investigation of various physical properties of $Hf_3PB_4$ will be useful for future experimental and theoretical research work on this interesting MAX phase nanolaminate.

# Recently predicted ternary boride $Hf_3PB_4$: Insights into the physical properties of this hardest possible boride MAX phase

M. A. Ali[a,1], M. M. Hossain[a], A. K. M. A. Islam[b,c], S. H. Naqib[c,2]

[a]Department of Physics, Chittagong University of Engineering and Technology (CUET), Chattogram4349, Bangladesh
[b]Department of Electrical and Electronic Engineering, International Islamic University Chittagong, Kumira, Chittagong, 4318, Bangladesh
[c]Department of Physics, University of Rajshahi, Rajshahi-6205, Bangladesh

**Table 1** Stiffness constants, $C_{ij}$ (GPa), bulk modulus, $B$ (GPa), shear modulus, $G$ (GPa), Young's modulus, $Y$ (GPa), Hardness parameters (GPa), and melting temperature, $T_m$ (K) of 211 MAX phases.

| Phase | $C_{11}$ | $C_{12}$ | $C_{13}$ | $C_{33}$ | $C_{44}$ | $B$ | $G$ | $Y$ | $*H_{macro}$ | $*H_{micro}$ | $*T_m$ |
|---|---|---|---|---|---|---|---|---|---|---|---|
| [1]$Ti_2AlC$ | 302 | 62 | 58 | 270 | 109 | 137 | 114 | 267 | 22.76 | 24.69 | 1665.00 |
| [1]$Ti_2AlN$ | 309 | 67 | 90 | 282 | 125 | 155 | 118 | 281 | 20.69 | 23.77 | 1704.00 |
| [1]$Ti_2GaC$ | 303 | 66 | 63 | 263 | 101 | 139 | 109 | 260 | 20.41 | 22.65 | 1657.50 |
| [2]$Ti_2SnC$ | 337 | 86 | 102 | 329 | 169 | 176 | 138 | 329 | 23.87 | 28.66 | 1858.50 |
| [3]$Ti_2SC$ | 368 | 108 | 123 | 395 | 189 | 204 | 151 | 363 | 23.48 | 29.85 | 2050.50 |
| [4]$Ti_2InC$ | 283 | 70 | 55 | 233 | 58 | 125 | 82 | 201 | 13.08 | 14.65 | 1552.50 |
| [5]$Ti_2GeC$ | 279 | 99 | 95 | 283 | 125 | 158 | 104 | 255 | 15.56 | 18.65 | 1615.50 |
| [6]$Ti_2CdC$ | 258 | 68 | 46 | 205 | 33 | 116 | 70 | 174 | 10.30 | 11.67 | 1435.50 |
| [7]$Ti_2TlC$ | 315 | 90 | 65 | 266 | 84 | 147 | 100 | 245 | 15.85 | 18.52 | 1698.00 |
| [8]$Ti_2AlB$ | 196 | 87 | 56 | 209 | 75 | 112 | 67 | 167 | 9.83 | 11.10 | 1255.50 |
| [9]$Ti_2AlB$ | 234 | 74 | 81 | 262 | 115 | 134 | 94 | 227 | 15.84 | 17.69 | 1449.00 |
| [9]$Ti_2GaB$ | 210 | 81 | 58 | 206 | 72 | 113 | 70 | 175 | 10.71 | 12.05 | 1293.00 |
| [9]$Ti_2InB$ | 203 | 73 | 49 | 190 | 59 | 104 | 65 | 160 | 10.27 | 11.11 | 1248.00 |
| [10]$Ti_2SiB$ | 250 | 75 | 81 | 263 | 120 | 137 | 99 | 240 | 17.11 | 19.27 | 1498.50 |
| [11]$Ti_2PbC$ | 239 | 94 | 49 | 21 | 69 | 120 | 76 | 188 | 11.76 | 13.23 | 1102.50 |
| [1]$Zr_2AlC$ | 261 | 63 | 63 | 224 | 87 | 125 | 92 | 221 | 16.68 | 18.07 | 1473.00 |
| [1]$Zr_2AlN$ | 264 | 77 | 89 | 235 | 105 | 141 | 94 | 231 | 14.75 | 17.11 | 1498.50 |
| [2]$Zr_2SnC$ | 269 | 94 | 81 | 290 | 148 | 157 | 110 | 268 | 17.63 | 20.86 | 1596.00 |
| [3]$Zr_2SC$ | 326 | 103 | 119 | 351 | 160 | 186 | 128 | 313 | 19.07 | 23.93 | 1858.50 |
| [12]$Zr_2SiC$ | 254 | 80 | 104 | 274 | 121 | 150 | 97 | 239 | 14.45 | 17.17 | 1527.00 |
| [12]$Zr_2PC$ | 293 | 90 | 113 | 343 | 145 | 173 | 117 | 286 | 17.52 | 21.49 | 1747.50 |
| [13]$Zr_2CdC$ | 278 | 77 | 78 | 287 | 100 | 146 | 101 | 245 | 16.33 | 18.83 | 1618.50 |
| [14]$Zr_2InC$ | 279 | 66 | 75 | 255 | 94 | 137 | 99 | 239 | 17.11 | 19.19 | 1573.50 |
| [5]$Zr_2GeC$ | 224 | 105 | 108 | 243 | 99 | 148 | 74 | 190 | 8.02 | 10.56 | 1390.50 |
| [11]$Zr_2PbC$ | 217 | 73 | 71 | 227 | 59 | 121 | 67 | 169 | 8.72 | 10.40 | 1345.50 |
| [7]$Zr_2TlC$ | 279 | 68 | 69 | 245 | 81 | 135 | 93 | 226 | 15.33 | 17.30 | 1558.50 |
| [8]$Zr_2AlB$ | 177 | 70 | 51 | 176 | 52 | 97 | 55 | 139 | 7.74 | 8.76 | 1149.00 |
| [9]$Zr_2AlB$ | 174 | 72 | 51 | 167 | 53 | 96 | 54 | 136 | 7.52 | 8.50 | 1126.50 |
| [9]$Zr_2GaB$ | 178 | 70 | 65 | 128 | 51 | 97 | 50 | 127 | 6.08 | 7.27 | 1080.00 |
| [9]$Zr_2InB$ | 184 | 62 | 46 | 174 | 45 | 94 | 54 | 139 | 7.79 | 8.87 | 1167.00 |
| [15]$Zr_2SB$ | 261 | 79 | 80 | 282 | 117 | 142 | 102 | 247 | 17.32 | 19.71 | 1560.00 |
| [1]$Sc_2AlC$ | 175 | 59 | 33 | 191 | 44 | 88 | 57 | 140 | 9.81 | 10.08 | 1165.50 |
| [1]$Sc_2GaN$ | 214 | 60 | 57 | 214 | 70 | 110 | 75 | 183 | 12.97 | 13.86 | 1317.00 |
| [1]$Sc_2InC$ | 175 | 59 | 33 | 173 | 41 | 86 | 54 | 135 | 8.97 | 9.42 | 1138.50 |
| [1]$Sc_2TlC$ | 180 | 54 | 30 | 166 | 37 | 84 | 55 | 135 | 9.70 | 9.82 | 1143.00 |
| [8]$Sc_2AlN$ | 210 | 70 | 54 | 218 | 71 | 110 | 73 | 179 | 12.23 | 13.20 | 1311.00 |
| [2]$Hf_2SnC$ | 330 | 54 | 126 | 292 | 167 | 173 | 132 | 316 | 22.36 | 26.79 | 1782.00 |
| [3]$Hf_2SC$ | 344 | 116 | 138 | 369 | 175 | 205 | 133 | 329 | 18.07 | 23.72 | 1939.50 |

*Continued..... Table 1*

| Phase | $C_{11}$ | $C_{12}$ | $C_{13}$ | $C_{33}$ | $C_{44}$ | $B$ | $G$ | $Y$ | $*H_{macro}$ | $*H_{micro}$ | $*T_m$ |
|---|---|---|---|---|---|---|---|---|---|---|---|
| [13]Hf2CdC | 290 | 108 | 84 | 307 | 37 | 160 | 66 | 173 | 5.23 | 7.93 | 1684.50 |
| [16]$Hf_2AlC$ | 317 | 74 | 80 | 266 | 118 | 152 | 116 | 277 | 20.52 | 23.52 | 1704.00 |
| [16]$Hf_2AlN$ | 329 | 95 | 122 | 301 | 144 | 182 | 121 | 297 | 17.51 | 21.95 | 1792.50 |
| [16]$Hf_2AlB$ | 199 | 78 | 60 | 190 | 76 | 109 | 68 | 169 | 10.59 | 11.71 | 1236.00 |
| [16]$Hf_2SiC$ | 306 | 86 | 116 | 305 | 146 | 172 | 118 | 288 | 17.97 | 21.96 | 1729.50 |
| [16]$Hf_2SiN$ | 250 | 120 | 157 | 299 | 148 | 183 | 88 | 228 | 8.66 | 12.16 | 1552.50 |
| [16]$Hf_2SiB$ | 223 | 95 | 102 | 238 | 120 | 142 | 83 | 208 | 11.15 | 13.53 | 1380.00 |
| [16]$Hf_2PC$ | 313 | 100 | 131 | 368 | 168 | 190 | 127 | 312 | 18.24 | 23.14 | 1845.00 |
| [16]$Hf_2PN$ | 319 | 115 | 146 | 367 | 173 | 201 | 125 | 311 | 16.34 | 21.46 | 1861.50 |
| [16]$Hf_2PB$ | 254 | 91 | 120 | 330 | 145 | 164 | 104 | 258 | 14.76 | 18.15 | 1611.00 |
| [14]$Hf_2InC$ | 331 | 87 | 90 | 284 | 101 | 168 | 109 | 270 | 15.75 | 19.46 | 1773.00 |
| [5]$Hf_2GeC$ | 269 | 96 | 125 | 278 | 128 | 167 | 97 | 244 | 12.39 | 15.75 | 1578.00 |
| [17]$Hf_2SnN$ | 266 | 99 | 138 | 268 | 110 | 171 | 84 | 216 | 8.63 | 11.79 | 1554.00 |
| [11]$Hf_2PbC$ | 245 | 73 | 70 | 230 | 76 | 128 | 81 | 201 | 12.31 | 14.13 | 1434.00 |
| [7]$Hf_2TlC$ | 307 | 84 | 80 | 267 | 83 | 152 | 97 | 240 | 14.18 | 17.02 | 1675.50 |
| [8]$Hf_2AlB$ | 204 | 76 | 58 | 200 | 68 | 110 | 67 | 168 | 10.10 | 11.37 | 1266.00 |
| [9]$Hf_2AlB$ | 232 | 72 | 81 | 267 | 109 | 133 | 92 | 223 | 15.31 | 17.14 | 1450.50 |
| [9]$Hf_2GaB$ | 213 | 77 | 63 | 176 | 66 | 111 | 67 | 167 | 9.96 | 11.20 | 1257.00 |
| [9]$Hf_2InB$ | 210 | 69 | 50 | 197 | 55 | 106 | 65 | 163 | 9.97 | 11.11 | 1279.50 |
| [15]$Hf_2SB$ | 286 | 79 | 84 | 296 | 122 | 151 | 111 | 267 | 18.94 | 21.81 | 1656.00 |
| [1]$V_2AlC$ | 339 | 71 | 100 | 319 | 148 | 171 | 134 | 319 | 23.39 | 27.78 | 1849.50 |
| [1]$V_2GaC$ | 334 | 81 | 111 | 299 | 138 | 175 | 125 | 302 | 19.74 | 23.97 | 1804.50 |
| [18]$V_2SnC$ | 243 | 76 | 124 | 300 | 87 | 156 | 82 | 209 | 9.41 | 12.21 | 1533.00 |
| [5]$V_2GeC$ | 282 | 121 | 144 | 259 | 160 | 182 | 99.5 | 253 | 11.55 | 15.37 | 1588.50 |
| [8]$V_2AlN$ | 302 | 45 | 131 | 332 | 143 | 169 | 121 | 294 | 19.37 | 23.39 | 1758.00 |
| [8]$V_2AlB$ | 285 | 86 | 98 | 278 | 139 | 157 | 111 | 270 | 17.96 | 21.21 | 1626.00 |
| [1]$Nb_2AlC$ | 310 | 90 | 118 | 289 | 139 | 173 | 116 | 285 | 17.21 | 21.23 | 1717.50 |
| [1]$Nb_2GaC$ | 309 | 80 | 138 | 262 | 126 | 177 | 108 | 270 | 14.36 | 18.31 | 1674.00 |
| [2]$Nb_2SnC$ | 341 | 106 | 169 | 321 | 183 | 209 | 126 | 315 | 15.73 | 21.10 | 1858.50 |
| [4]$Nb_2InC$ | 291 | 77 | 118 | 289 | 57 | 182 | 80 | 209 | 6.92 | 10.21 | 1660.50 |
| [4]$Nb_2AsC$ | 325 | 114 | 161 | 326 | 150 | 234 | 117 | 301 | 11.41 | 16.72 | 1818.00 |
| [4]$Nb_2SC$ | 304 | 117 | 155 | 316 | 88 | 221 | 89 | 234 | 6.53 | 10.47 | 1740.00 |
| [5]$Nb_2GeC$ | 308 | 133 | 168 | 306 | 177 | 206 | 109 | 279 | 11.77 | 16.40 | 1737.00 |
| [8]$Nb_2AlN$ | 340 | 150 | 111 | 332 | 148 | 195 | 119 | 296 | 15.38 | 20.07 | 1872.00 |
| [8]$Nb_2AlB$ | 284 | 91 | 102 | 249 | 125 | 156 | 102 | 252 | 15.20 | 18.31 | 1579.50 |
| [19]$Nb_2SB$ | 316 | 95 | 131 | 317 | 143 | 186 | 116 | 287 | 15.57 | 19.89 | 1777.50 |
| [1]$Ta_2AlC$ | 334 | 114 | 130 | 322 | 148 | 193 | 122 | 303 | 16.43 | 21.28 | 1839.00 |
| [1]$Ta_2GaC$ | 335 | 106 | 137 | 315 | 137 | 194 | 118 | 294 | 15.22 | 19.87 | 1831.50 |
| [13]$Ta_2CdC$ | 408 | 106 | 108 | 218 | 151 | 179 | 133 | 320 | 21.69 | 26.42 | 1905.00 |
| [20]$Ta_2InC$ | 396 | 102 | 133 | 345 | 133 | 208 | 133 | 329 | 17.71 | 23.37 | 2059.50 |
| [5]$Ta_2GeC$ | 370 | 147 | 194 | 389 | 220** | 243 | 140 | 352 | 15.89 | 22.50 | 2047.50 |
| [8]$Ta_2AlN$ | 352 | 174 | 143 | 373 | 160 | 222 | 119 | 302 | 12.79 | 17.99 | 1969.50 |
| [8]$Ta_2AlB$ | 325 | 97 | 113 | 276 | 141 | 174 | 117 | 288 | 17.38 | 21.52 | 1743.00 |
| [1]$Cr_2AlC$ | 365 | 84 | 102 | 369 | 140 | 186 | 138 | 332 | 22.19 | 27.37 | 2002.50 |
| [5]$Cr_2GeC$ | 315 | 148 | 146 | 354 | 89 | 207 | 88 | 232 | 7.09 | 10.96 | 1830.00 |
| [8]$Cr_2AlN$ | 287 | 75 | 145 | 377 | 88 | 181 | 95 | 242 | 10.50 | 14.11 | 1780.50 |
| [8]$Cr_2AlB$ | 299 | 82 | 130 | 290 | 157 | 174 | 116 | 285 | 17.08 | 21.11 | 1686.00 |
| [1]$Mo_2AlC$ | 333 | 97 | 144 | 327 | 137 | 205 | 127 | 296 | 16.43 | 20.38 | 1843.50 |
| [4]$Mo_2GaC$ | 306 | 105 | 169 | 311 | 102 | 249 | 96 | 254 | 6.47 | 10.88 | 1738.50 |
| [5]$Mo_2GeC$ | 331 | 136 | 184 | 342 | 123 | 223 | 100 | 260 | 8.57 | 12.95 | 1860.00 |
| [8]$Mo_2AlB$ | 317 | 105 | 152 | 269 | 161 | 191 | 111 | 279 | 13.66 | 18.02 | 1708.50 |
| [5]$W_2GeC$ | 340 | 146 | 222 | 368 | 117 | 244 | 93 | 247 | 6.17 | 10.46 | 1926.00 |
| [21]$Lu_2SnC$ | 172 | 46 | 36 | 173 | 56 | 82 | 61 | 147 | 12.67 | 12.15 | 1129.50 |

*Calculated using published data.

** This value was calculated using local density approximation (LDA) that overestimates the elastic constants. We have calculated the same parameter as 178 GPa using GGA with identical cut-off energy and *k*-points grid size of Ref.[5]. It is noteworthy that the calculated value of $C_{44}$ for $Hf_3PB_4$ using LDA is 244 GPa.

**Table 2:** Stiffness constants, $C_{ij}$ (GPa), bulk modulus, $B$ (GPa), shear modulus, $G$ (GPa), Young's modulus, $Y$ (GPa), Hardness parameters (GPa), and melting temperature, $T_m$ (K) of 312 MAX phases.

| Phase | $C_{11}$ | $C_{12}$ | $C_{13}$ | $C_{33}$ | $C_{44}$ | $B$ | $G$ | $Y$ | *$H_{macro}$ | *$H_{micro}$ | *$T_m$ |
|---|---|---|---|---|---|---|---|---|---|---|---|
| [22]$Ti_3AlC_2$ | 368 | 81 | 76 | 313 | 130 | 168 | 135 | 320 | 24.30 | 28.57 | 1927.50 |
| [23]$Ti_3GaC_2$ | 359 | 78 | 69 | 294 | 123 | 159 | 130 | 306 | 18.77 | 21.66 | 1768.50 |
| [24]$Ti_3GeC_2$ | 355 | 85 | 94 | 338 | 148 | 177 | 138 | 312 | 24.25 | 27.80 | 1872.00 |
| [22]$Ti_3InC_2$ | 340 | 85 | 67 | 263 | 97 | 152 | 111 | 267 | 23.69 | 27.03 | 1926.00 |
| [24]$Ti_3SiC_2$ | 372 | 88 | 98 | 353 | 167 | 185 | 149 | 352 | 26.00 | 31.50 | 1999.50 |
| [24]$Ti_3SnC_2$ | 331 | 91 | 81 | 299 | 129 | 162 | 122 | 285 | 20.85 | 23.85 | 1795.50 |
| [25]$Zr_3AlC_2$ | 322 | 84 | 97 | 287 | 138 | 165 | 122 | 294 | 20.34 | 24.12 | 1750.50 |
| [25]$Zr_3SiC_2$ | 323 | 85 | 99 | 303 | 135 | 169 | 122 | 295 | 19.70 | 23.66 | 1777.50 |
| [26]$Zr_3SnC_2$ | 280 | 92 | 84 | 257 | 110 | 148 | 99 | 243 | 15.37 | 18.06 | 1579.50 |
| [27]$Hf_3AlC_2$ | 347 | 77 | 80 | 291 | 127 | 162 | 127 | 302 | 22.59 | 26.31 | 1831.50 |
| [22]$Hf_3SiC_2$ | 348 | 101 | 120 | 335 | 144 | 190 | 127 | 312 | 18.24 | 23.14 | 1900.50 |
| [22]$Hf_3SnC_2$ | 326 | 96 | 97 | 300 | 107 | 170 | 110 | 271 | 15.80 | 19.52 | 1782.00 |
| [25]$V_3AlC_2$ | 404 | 84 | 108 | 361 | 158 | 197 | 153 | 364 | 25.23 | 31.41 | 2107.50 |
| [22]$Ta_3AlC_2$ | 441 | 132 | 138 | 382 | 175 | 231 | 157 | 384 | 21.51 | 29.00 | 2250.00 |
| [22]$Ta_3SiC_2$ | 352 | 220 | 210 | 345 | 182 | 256 | 102 | 270 | 07.20 | 11.95 | 1927.50 |
| [25]$Mo_3SiC_2$ | 377 | 175 | 186 | 364 | 151 | 245 | 116 | 301 | 10.45 | 15.83 | 2031.00 |

*Calculated using published data.

**Table 3:** Stiffness constants, $C_{ij}$ (GPa), bulk modulus, $B$ (GPa), shear modulus, $G$ (GPa), Young's modulus, $Y$ (GPa), Hardness parameters (GPa), and melting temperature, $T_m$ (K) of 413 MAX phases.

| Phase | $C_{11}$ | $C_{12}$ | $C_{13}$ | $C_{33}$ | $C_{44}$ | $B$ | $G$ | $Y$ | *$H_{macro}$ | *$H_{micro}$ | *$T_m$ |
|---|---|---|---|---|---|---|---|---|---|---|---|
| [28]$Ti_4AlN_3$ | 407 | 95 | 100 | 364 | 162 | 196 | 155 | 367 | 26.05 | 32.25 | 2121.00 |
| [29]$Ti_4GaC_3$ | 353 | 68 | 57 | 276 | 130 | 150 | 134 | 332 | 27.77 | 32.95 | 1827.00 |
| [29]$Ti_4GeC_3$ | 355 | 74 | 80 | 321 | 138 | 167 | 137 | 327 | 25.21 | 29.81 | 1900.50 |
| [30]$Ti_4SiC_3$ | 398 | 173 | 96 | 378 | 173 | 197 | 157 | 357 | 26.53 | 31.61 | 2115.00 |
| [31]$V_4AlC_3$ | 435 | 121 | 105 | 384 | 168 | 218 | 170 | 414 | 27.16 | 35.87 | 2235.00 |
| [30]$V_4SiC_3$ | 415 | 155 | 150 | 410 | 174 | 239 | 146 | 333 | 17.74 | 22.60 | 2214.00 |
| [29]$Nb_4AlC_3$ | 428 | 109 | 123 | 355 | 168 | 213 | 156 | 374 | 23.65 | 30.44 | 2170.50 |
| [29]$Nb_4SiC_3$ | 388 | 149 | 170 | 356 | 186 | 235 | 136 | 342 | 15.67 | 21.99 | 2052.00 |
| [29]$Nb_4GaC_3$ | 424 | 123 | 136 | 346 | 155 | 220 | 154 | 375 | 22.09 | 29.17 | 2145.00 |
| [29]$Nb_4GeC_3$ | 404 | 149 | 153 | 319 | 161 | 225 | 132 | 332 | 15.65 | 21.64 | 2044.50 |
| [32]$Ta_4AlC_3$ | 454 | 157 | 156 | 376 | 201 | 247 | 161 | 397 | 20.69 | 28.75 | 2280.00 |
| [33]$Ta_4SiC_3$ | 396 | 190 | 180 | 391 | 207 | 254 | 138 | 350 | 14.49 | 21.13 | 2128.50 |

*Calculated using published data.

**Table 4:** Stiffness constants, $C_{ij}$ (GPa), bulk modulus, $B$ (GPa), shear modulus, $G$ (GPa), Young's modulus, $Y$ (GPa), Hardness parameters (GPa), and melting temperature, $T_m$ (K) of 514 MAX phases.

| Phase | $C_{11}$ | $C_{12}$ | $C_{13}$ | $C_{33}$ | $C_{44}$ | $B$ | $G$ | $Y$ | *$H_{macro}$ | *$H_{micro}$ | *$T_m$ |
|---|---|---|---|---|---|---|---|---|---|---|---|
| [34]$Ti_5AlC_4$ | 381 | 82 | 76 | 320 | 134 | 172 | 140 | 330 | 25.31 | 29.84 | 1977.00 |
| [35]$Ti_5SiC_4$ | 418 | 92 | 108 | 390 | 176 | 205 | 163 | 387 | 27.11 | 34.19 | 2193.00 |

*Calculated using published data.

**Table 5:** Debye temperature, $\Theta_D$ (K) of Hf-based MAX phases.

| Phase | $\Theta_D$ (K) |
|---|---|
| [7]$Hf_2TlC$ | 339 |
| [2]$Hf_2SnC$ | 393 |
| [20]$Hf_2InC$ | 383 |
| [5]$Hf_2GeC$ | 382 |
| [3]$Hf_2SC$ | 463 |
| [8]$Hf_2AlB$ | 342 |
| [8]$Hf_2AlC$ | 415 |
| [8]$Hf_2AlN$ | 427 |
| [15]$Hf_2SB$ | 430* |
| [16]$Hf_2SiC$ | 435* |
| [17]$Hf_2PC$ | 449* |
| [17]$Hf_2SiB$ | 369* |
| [17]$Hf_2PN$ | 443* |
| [17]$Hf_2SiN$ | 375* |
| [27]$Hf_3AlC_2$ | 459 |
| [26]$Hf_3SnC_2$ | 465 |

*Calculated using published data.